\documentclass[11pt,a4paper]{article}
\usepackage[T1]{fontenc} 
\usepackage[utf8]{inputenc}
\usepackage[a4paper, total={7in, 10in}]{geometry}
\usepackage{authblk}
\usepackage{graphicx}
\graphicspath{{Figures/}}
\usepackage[tbtags]{amsmath}
\usepackage{amssymb,mathtools,physics}
\usepackage{acronym}
\usepackage[labelfont=bf,format=plain, font=small]{caption}
\usepackage[square, comma, numbers, sort&compress]{natbib}
\usepackage{hyperref}
\hypersetup{
	colorlinks=false, breaklinks=true,
	urlcolor=webbrown, linkcolor=RoyalBlue, citecolor=webgreen,
	pdftitle={Effect of Quantum Gravity on Specific Heat of Solid},
	pdfauthor={Sheikh Riasat,Bhabani Prasad Mandal}, 
	pdfsubject={generalized uncertainty principle}, 
	pdfkeywords={gup, hep, einstein, debye, quantum, gravity},
	pdfcreator={pdfLaTeX}
}
\usepackage{cleveref}

\newcommand{\ie}{i.\,e.}

\newacro{dsr}[DSR]{doubly special relativity}
\newacro{gup}[GUP]{generalized uncertainty principle}
\newacro{hup}[HUP]{Heisenberg uncertainty principle}
\newacro{stm}[STM]{scanning tunneling microscope}
\newacro{sho}[SHO]{simple harmonic oscillator}
\newacro{qho}[QHO]{quantum harmonic oscillator}
\newacro{cho}[CHO]{coupled harmonic oscillator}

\begin{document}

\title{\bf{Effect of Quantum Gravity on Specific Heat of Solid}}
\author[a]{Sheikh Riasat\thanks{riasat.sheikh@icloud.com}}
\author[b]{Bhabani Prasad Mandal\thanks{bhabani.mandal@gmail.com}}
\affil[a, b]{\small{\em Department of Physics, Institute of Science, Banaras Hindu University, Varanasi, India-221005.}}
\date{July 15, 2022}
\maketitle

\begin{abstract}
	All possible theories of  quantum gravity suggest the existence of a minimal length. As a consequence, the usual \ac{hup} is replaced by a more general uncertainty principle  known as the \ac{gup}. The dynamics of all quantum mechanical system gets modified due to GUP. In this work, we consider both Einstein's and Debye's models to find the quantum gravity effect on the specific heat of solids. GUP modified specific heat in Einstein's model  shows usual exponential dominance at low temperatures. Further, the modification to Debye's specific heat is calculated by considering the GUP modified dispersion relation, which becomes time dependent for elastic waves.
\end{abstract}

\section{Introduction}\label{sec:1}

Three of the fundamental interactions are well described in the quantum domain whereas gravitational interaction, the weakest one,  is understood in the classical domain using the general theory of relativity. However, gravity becomes important at very high energy  and hence its effects can not be ignored at short distances. Thus one of the major motivations is to develop a consistent and complete theory of gravity by reconciling gravity and quantum mechanics in the same framework. However, the formulation of a quantum theory of gravity is an ongoing challenge for theoretical physics. There exists a lot of promising candidates \cite{tawfik-Generalized-2014} of the same such as string theory \cite{amati-Can-1989}, doubly special relativity \cite{amelino-camelia-Doubly-2002}, black hole physics \cite{scardigli-Generalized-1999}, canonical quantum gravity, etc.. All of these theories  predict the existence of a minimal length \cite{amati-Can-1989,gross-HighEnergy-1987,gross-String-1988,tawfik-Generalized-2014,konishi-Minimum-1990,ali-Discreteness-2009,ali-Proposal-2011} \ie, the Planck length $(l_{_{pl}}\approx 10^{-35} \,m)$. The mere existence of a minimum approachable length compels us to generalize the \ac{hup}. Out of the numerous formulations of \ac{gup} \cite{tawfik-Generalized-2014,kempf-Hilbert-1995,ali-Discreteness-2009,kempf-Nonpointlike-1997}, the most general one is a quadratic momentum-dependent uncertainty relation \cite{kempf-Hilbert-1995} given as,
\begin{equation}\label{eq:1}
	\Delta q \Delta p \geq \frac{1}{2}\qty(1 + \gamma^2 \qty(\Delta p)^2 + \gamma^2 \ev{p}^2)
\end{equation}
where,
\begin{equation}\label{eq:2}
	\gamma = \frac{\gamma_{_0}}{M_{_{pl}}c}
\end{equation}
is a positive quantity and $\gamma_{_0}$ is the dimensionless \ac{gup} parameter which can be determined experimentally. This parameter tells us about the working length scale.  We can estimate this by finding an upper bound on $\gamma_{_0}$, which tending unity brings the length scale closer to Planck scale \cite{ali-Proposal-2011,ali-Discreteness-2009} . The commutation relation in one dimension which is consistent with  \cref{eq:1} is then modified as,
\begin{equation}\label{eq:3}
	\comm{q}{p} =i \hbar\qty(1 + \gamma^2 p^2)
\end{equation}
The above \ac{gup} modified commutation relation for the  position $q$ and momentum $p$ operators are written  in terms of the low-energy quantities, $q_0$ and $p_0$ respectively \cite{bosso-Planck-2017} \ie,
\begin{subequations}
	\begin{align}
		q & = q_0                                    \\
		p & = p_0 \qty(1 +  \frac{\gamma^2}{3}p_0^2)
	\end{align}
\end{subequations}
where\qcomma $\comm{q_0}{p_0} = i \hbar$.
The effect due to presence of minimum length in a quantum mechanical system is then obtained by  substituting  the above high energy momentum in the   Hamiltonian of  that system. For example the quantum gravity effect on a simple harmonic oscillator up to $\order{\gamma^2}$ is obtained by studying
\begin{equation}\label{eq:5}
	H = \frac{p_0^2}{2m} + \frac{1}{2}m\omega^2 q_0^2 + \frac{\gamma^2}{3m}p_0^4
\end{equation}
In the limit $\gamma \to 0$ we get back our standard hamiltonian.

Significant amount of research works have been  done to observe the effect of gravity in various branches of physics including Black hole physics \cite{Scardigli_1999,GECIM2017391,_vg_n_2017, MAZIASHVILI2006232,Chen_2014}, Cosmology \cite{majumder2011effects} and quantum mechanical problems such as Landau Levels, Lamb Shift, potential barrier  \cite{das-Phenomenological-2009,ali-Proposal-2011}, \ac{sho} \cite{ali-Proposal-2011}, coherent and squeezed states \cite{bosso-Planck-2017} particle in a box \cite{ali-Discreteness-2009,ali-Proposal-2011}, angular momentum algebra, CG coefficients \cite{bosso-Generalized-2017,verma-Schwinger-2018},  non-Hermitian interactions \cite{Faizal_2015,Bagchi_2009}. Recently many quantum statistical mechnical systems have been studied in GUP framework \cite {Nouicer_2007, El-Nabulsi:2020zyh,Hamil:2020ldl, bouaziz, abhishek2023effect}. The effect of \ac{gup} has also been investigated in the relativistic limit such as Dirac oscillators \cite{mandal-Dirac-2010,yang-Dirac-2020,tyagi-gup-2020} and Klein-Gordon equations \cite{das-Discreteness-2010a}. Furthermore, prospective experimental tests have been suggested considering microscopic harmonic oscillators\cite{bawaj-Probing-2015},   macroscopic variables  \cite{marin-Gravitational-2013}  and  using quantum optomechanics \cite{pikovski-Probing-2012,bosso-Amplified-2017}

Recently, \ac{gup} modified energy spectrum of harmonic oscillator \cite{bosso-Minimal-2021,bosso-Planck-2017} was calculated and used to study the Planck's distribution law and thermodynamics of the black body radiation by calculating the Wein's law and Stefan-Boltzmann law \cite{bosso-Minimal-2021}. In this  work, \ac{gup} modification to the electromagnetic field quantization is considered by using
\begin{equation}
	\comm{\vb{q_k}}{\vb{p_{k'}}} = i\hbar\delta_{kk'}[1+\gamma_{_{EM}}^2 \vb{p^2_k}]
\end{equation}
where\qcomma $\vb{q_k}$  and $\vb{p_k}$ represent generalized coordinates and momenta of the electromagnetic field. Here the dimension of $\vb{p_k}$ is momentum/(mass)$^{1/2}$ and that of $\vb{q_k}$ is position$\times$(mass)$^{1/2}$, and
\begin{equation}
	\gamma_{_{EM}}=\frac{\gamma_{_0}}{\sqrt{M_{_{pl}}}c}
\end{equation} \cite{bosso-Potential-2018}.
Then the hamiltonian of \cref{eq:5} is written as,
\begin{equation}\label{eq:8}
	H = \frac{p_k^2}{2} + \frac{1}{2}\omega^2 q_k^2 + \frac{1}{3}\gamma_{_{EM}}^2 p_k^4
\end{equation}
Using this modification, \ac{gup} modified  energy spectrum  for a mode with wave vector $\vb{k}$ is calculated as,
\begin{equation}\label{eq:9}
	E_n^{k,GUP} = \hbar\omega^{k}\qty{\qty(n + \frac{1}{2}) + \frac{\hbar\omega^k}{4}\gamma_{_{EM}}^2 (1+2n+2n^{2})}
\end{equation}
where\qcomma $k$ is the magnitude of the wave number used to label the different modes. In the limit $\gamma_{_{EM}}^2  \to 0$ we recover the usual expression for the energy of photons.

Einstein  fixed the discrepancy of Dulong-Petit law  for the specific heat of solids at low temperatures by considering the thermal vibrations of solid as independent \ac{qho} which vibrates at a constant frequency and posses discrete energy values. While Einstein's model was able to answer the problem in hand, it was  not completely agreeing with the exact experimental results. Finally  Debye improved it  by considering  \ac{cho} modes for the thermal vibrations. In this article we intend to study both the specific heat models in the framework of quantum mechanics with GUP  to obtain the quantum gravity effects  in the specific heat of solid.

This paper is organized as follows. In \cref{sec:2}, we consider the \ac{gup} Einstein's specific heat model and calculate the modification to the specific heat.  Debye model of specific heat is considered in  \cref{sec:3} in GUP framework. There we first derive the GUP modification to the dispersion relation of elastic waves  and use that  along with  the modified average energy of the oscillators to find how quantum gravity  affect the specific heat in the \ac{gup} modified Debye's specific heat model. \Cref{sec:4} is devoted to concluding remarks.

\section{Modified Einstein's Specific heat model}\label{sec:2}

The classical theory of specific heat  was unable to explain experimentally observed behaviour at low temperature.
Einstein considered the solid as collection of independent \acf{qho} with fixed frequency $\omega$.
Total energy $E(T)  $ due to  the thermal vibrations of  these \ac{qho} is then calculated by considering the average energy per oscillator instead of using classical law of equipartition of energy. In this section we revisit this path breaking formulation in  the modified quantum theory with GUP to see the quantum gravity effect in the expression of  specific heat of Solid. We calculate the average energy per QHO using modified energy spectrum of the
oscillators.

Now to calculate specific heat of solid in  the modified quantum theory with GUP we calculate the average energy per oscillator using the Boltzmann distribution. The probability that a single mode has energy $E_n^{^{GUP}}$ of \cref{eq:9} is given by the usual Boltzmann factor as
\begin{equation}
	\Pr(n) = \frac{\exp[-E_n^{^{GUP}}/k_{_B} T]}{\sum\limits_{n=0}^\infty \exp[-E_n^{^{GUP}}/k_{_B} T] }
\end{equation}
Therefore, the \ac{gup} modified mean energy of the oscillators is given as
\begin{equation}
	\bar{E}^{^{GUP}} = \sum\limits_{n=0}^\infty E_n^{^{GUP}} \Pr(n) = \frac{\sum\limits_{n=0}^\infty E_n^{^{GUP}} \exp[-E_n^{^{GUP}}/k_{_B} T]}{\sum\limits_{n=0}^\infty \exp[-E_n^{^{GUP}}/k_{_B} T] }
\end{equation}
Now, denoting $b = \hbar\omega\gamma_{_{EM}}^2$ and $x = - \hbar \omega / k_{_B} T$ and simplifying the above equation up to $\order{\gamma_{_{EM}}^2}$ we can have,
\begin{equation}\label{eq:12}
	\bar{E}^{^{GUP}} = \hbar \omega \qty(\frac{1}{2} + \frac{b}{4}) + \hbar \omega \dv{x}\ln(z)
\end{equation}
where,
\begin{equation}
	z = \sum\limits_{j=0}^\infty \exp[\qty(j + \frac{j(j+1)}{2}b)x]
\end{equation}
The second term in \cref{eq:12} is  calculated up to $\order{\gamma_{_{EM}}^2}$ as,
\begin{equation}\label{eq:14}
	\hbar \omega \dv{x} \ln(z) = \frac{\hbar \omega}{e^{-x} - 1} + b\hbar  \omega \sum\limits_{j=1}^\infty \qty(1 + xj)je^{jx}
\end{equation}
Using \cref{eq:12,eq:14},  the \ac{gup} modified mean energy of the \acp{qho} becomes,
\begin{equation}\label{eq:15}
	\bar{E}^{^{GUP}} = \bar{E} + \bar{E}\,'
\end{equation}
where,
\begin{subequations}
	\begin{align}
		\bar{E}    & = \hbar \omega \qty(\frac{1}{2} + \frac{1}{e^{\hbar \omega/ k_{_B} T} - 1})                                                                        \\
		\bar{E}\,' & = \hbar^2 \omega^2 \gamma_{_{EM}}^2 \qty(\frac{1}{4} + \sum\limits_{j=1}^\infty (1 - \frac{\hbar \omega}{k_{_B} T}j)je^{-j\hbar \omega/ k_{_B} T})
	\end{align}
\end{subequations}

The total internal energy of the solid under the modification of \ac{gup} is given as,
\begin{equation}
	U^{^{GUP}} = 3N\bar{E}^{^{GUP}}
\end{equation}
and therefore, the \ac{gup} modified specific heat of the solid in Einstein's formulation is
\begin{equation}
	C_v^{^{GUP}} = \dv{U^{^{GUP}}}{T} = C_{v} + C\,'_{v}
\end{equation}
where,
\begin{subequations}
	\begin{align}
		C_{v}    & = 3Nk_{_B} \frac{\theta_{_E}^2}{T^2} \frac{e^{\theta_{_E}/ T}}{\qty(e^{\theta_{_E}/ T} - 1)^2}                                               \\
		C\,'_{v} & = 3Nk_B^2 \gamma_{_{EM}}^2 \, \frac{\theta_{_E}^4}{T^3} \sum\limits_{j=1}^\infty \qty(\frac{2}{\theta_{_E}/ T} - j) j^2 e^{-j\theta_{_E}/ T}
	\end{align}
\end{subequations}
where\qcomma $\theta_{_E} = \hbar \omega/ k_{_B} $ is the characteristic temperature known as Einstein's temperature. In the limit $\gamma_{_{EM}}^2 \to 0$, $C\,'_{v} \to 0$ for solid.

\begin{figure}[ht]
	\centering
	\includegraphics[scale=0.4]{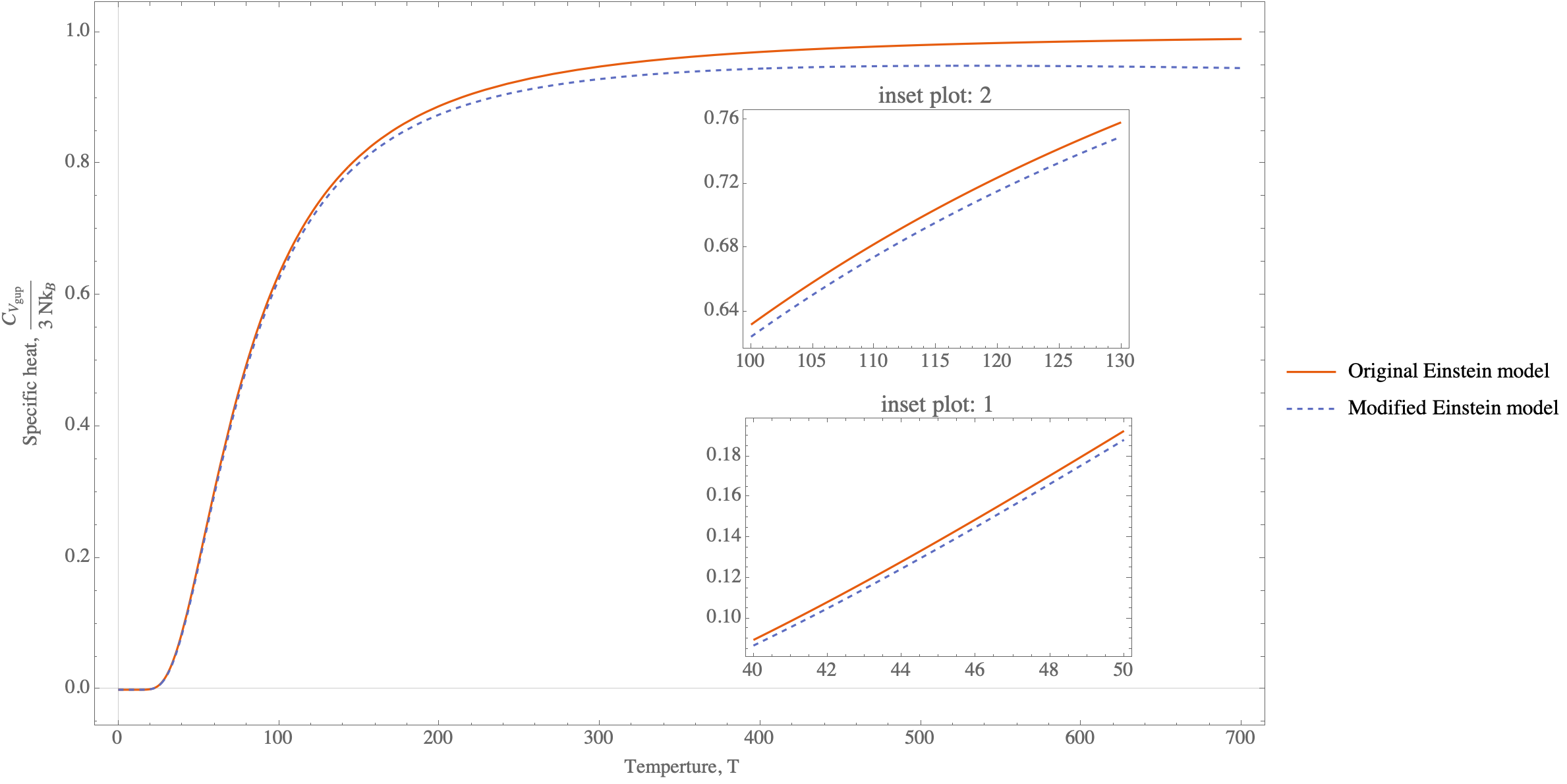}
	\caption[modified Einstein]{Modified Einstein specific heat, $C_v^{^{GUP}} / 3Nk_{_B}$ vs. Temperature, $T$ plot.  Here we have considered $\theta_E \approxeq 240K$ (for copper \cite{mahmood-Experimental-2011})  and the temperature range $0K \leq T\leq 700K$. We have chosen the \ac{gup} dependent parameter $k_{_B} \gamma_{_{EM}}^2 = 10^{-4.5}$. Inset plot $1$ and $2$ shows the variation of specific heat at $40K \leq T\leq 50K$ and $100K \leq T\leq 130K$ respectively.}
	\label{fig:1}
\end{figure}

We now consider high temperature and low temperature behaviors of this specific heat.
\paragraph{Behavior at high temperature:}  Let us denote $\delta = \theta_{_E}/ T$. For temperatures $T \gg \theta_E $ or $\delta \ll 1$ we can write $e^\delta \approxeq 1 + \delta $ and $e^{-j\delta} \approxeq 1 - j \delta $. Thus, at high temperature the \ac{gup} modified specific heat of the solid is given as\footnote{$\uparrow$  denotes high temperatures and $\downarrow$  denotes low temperatures}
\begin{equation}
	C_v^{^{GUP^{\uparrow}}} = C_{v}^{^\uparrow} + C\,'_{v}{^{^\uparrow}}
\end{equation}
where,
\begin{subequations}
	\begin{align}
		C_{v}^{^\uparrow}      & = 3Nk_{_B}\label{eq:21.a}                                                                                                                       \\
		C\,'_{v}{^{^\uparrow}} & = 3Nk_B^2 \gamma_{_{EM}}^2 \, \frac{\theta_{_E}^3}{T^2} \sum\limits_{j=1}^\infty \qty(2 - \frac{\theta_{_E}}{T}\qty(2j + 1)) j^2\label{eq:21.b}
	\end{align}
\end{subequations}
The original term \cref{eq:21.a}  reduces to the classical expression \ie, the Dulong-Petit law. The expression  in \cref{eq:21.b} represents the  GUP modification of   $C_v$  at high temperature  and additional have a negative value. So, the overall specific heat at high temperature tends to fall from the classical value as observed in \cref{fig:1}.

\begin{figure}[ht]
	\centering
	\includegraphics[scale=0.4]{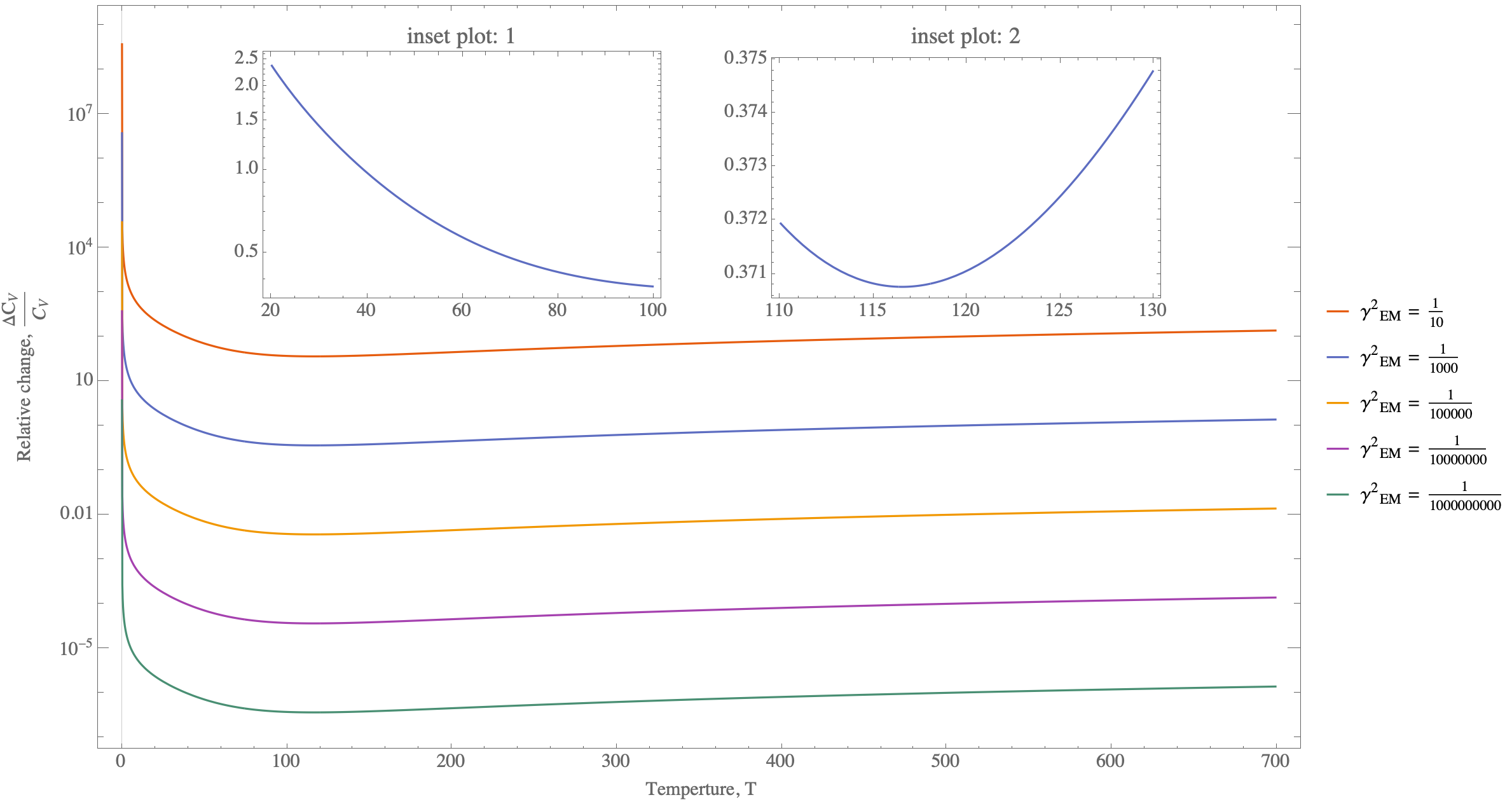}
	\caption[relative Einstein]{Relative Einstein specific heat, $\Delta C_v/ C_{v}$ vs. Temperature, $T$ plot at $\theta_E \approxeq 240K$ with temperature range $0K\leq T \leq 700K$ and varying the \ac{gup} parameter as $\gamma_{_{EM}}^2 = \qty{10^{-1}, 10^{-3}, 10^{-5}, 10^{-7}, 10^{-9}}$. Inset plot $1$ and $2$ shows the relative change of specific heat at $20K\leq T \leq 100K$ and $110K\leq T \leq 130K$ respectively for the \ac{gup} parameter $\gamma_{_{EM}}^2 = 10^{-3}$}
	\label{fig:2}
\end{figure}

\paragraph{Behavior at low temperature:} For temperatures $T \ll \theta_{_E}$ or $\delta \gg 1$ we can write $e^\delta - 1 \approxeq e^\delta$. Thus, at low temperature the \ac{gup} modified specific heat of the solid is given as
\begin{equation}
	C_v^{^{GUP^{\downarrow}}} = C_{v}^{^\downarrow} + C\,'_{v}{^{^\downarrow}}
\end{equation}
where,
\begin{subequations}
	\begin{align}
		C_{v}^{^\downarrow}      & = 3Nk_{_B} \frac{\theta_{_E}^2}{T^2} e^{-\theta_{_E}/ T}\label{eq:23.a}                                                      \\
		C\,'_{v}{^{^\downarrow}} & = -3Nk_{_B}^2 \gamma_{_{EM}}^2 \, \frac{\theta_{_E}^4}{T^3} \sum\limits_{j=1}^\infty j^3 e^{-j\theta_{_E}/ T}\label{eq:23.b}
	\end{align}
\end{subequations}
Thus at low temperatures, the exponential term plays a major role in determining the temperature variation of specific heat for both the original term \cref{eq:23.a} and the additional term introduced due to \ac{gup} \cref{eq:23.b}. But the overall specific heat of the solid at low temperatures will stay at a lower value than usual as observed in \cref{fig:1} because of the negative value of the \ac{gup} introduced term.

We can further study the \ac{gup} modification in the Einstein specific heat model by plotting the relative change \ie,
\begin{equation}\label{eq:24}
	\frac{\Delta C_v}{C_{v}}=\frac{C_{v} - C_v^{^{GUP}}}{C_{v}}
\end{equation}

In \cref{fig:2}, we observe the relative modification for different values of \ac{gup} parameter $\gamma_{_{EM}}^2$. As the \ac{gup} parameter $\gamma_{_{EM}}^2$ reduces, the difference between original specific heat and the \ac{gup} modified specific heat decreases. Furthermore, for a particular value of $\gamma_{_{EM}}^2$ below $115 K$ the difference in specific heat reduces and beyond that a growth is observed which is consistent with our earlier observation in \cref{fig:1}.

\section{Modified Debye's Specific heat model}\label{sec:3}
Debye further improved the formulation  by considering the fact that the atoms are bound together in a solid and which cannot vibrate independently. The motion of \acf{cho}  can be described in terms of normal modes. These normal modes  are then considered to calculate the specific heat. If the solid has $N$ atoms then there will be a total of $3N$ normal modes of vibration and these modes  have different frequencies up to a maximum frequency known as Debye frequency, $\omega_D$. Total internal energy of solid in this model is then given by
\begin{equation}\label{eq:25}
	U^{^{GUP}} = 3 \int \dd{\omega} D(\omega)^{^{GUP}} \bar{E}(\omega)^{^{GUP}}
\end{equation}
where $D(\omega) \dd{\omega}$ number of modes in the frequency range $\omega$ and $\omega + \dd{\omega}$. We revisit this formulation in the framework of GUP modified quantum theories. We first need to calculate the total number of possible modes and hence $D(\omega)_{_{GUP}}$ of vibration in the Debye frequency range.

If we apply periodic boundary conditions \cite{kittel-Introduction-2018} over $N$ primitive cells within a cube of side $L$ and volume $V = L^3$, the total number of modes within the range $k$ and $k + \dd{k}$ is given as $N = \frac{V}{2\pi^2} \int \dd{k} k^2$ which is a function of wavevector $k$. The \ac{gup} modified hamiltonian \cref{eq:8} is suggesting a modification in the  equation of motion and thus changing the dispersion relation of elastic modes. The Hamilton's equation  for the GUP modified system in \cref{eq:5} up to $\order{\gamma_{_{EM}}^2 }$  is given as,
\begin{equation}\label{eq:26}
	\dot{p}_k = \ddot{q}_k \qty(1 - 4\gamma_{_{EM}}^2 \dot{q}_k^2)
\end{equation}
The total force acting on the $s^{th}$ atom due to the neighboring atoms assuming all atoms form a mass spring system with constant spring constant $\beta$ is,
\begin{equation}
	F_s = \beta \qty(u_{s+1} + u_{s-1} -2u_s)
\end{equation}
where\qcomma $u_s$ is the displacement of the $s^{th}$ atom  from its equilibrium position. The \cref{eq:26}   for our system is then written as,
\begin{equation}\label{eq:28}
	\ddot{u}_s \qty(1 - 4\gamma_{_{EM}}^2 \dot{u}_s^2) =  F_s =  \beta \qty(u_{s+1} + u_{s-1} - 2u_s)
\end{equation}
where\qcomma $u_s \sim q_k$. If we consider a wave solution of the form,
\begin{equation}
	u_s = u_0 e^{i\qty(\omega t - ksa)}
\end{equation}
where\qcomma $sa$ represents the  equillibrium position of the $s^{th}$ atom in the plane and $u_0$ represents the amplitude of the wave. Substituting this in \cref{eq:28} and simplifying up to $\order{\gamma_{_{EM}}^2 }$ we get the \ac{gup} modified dispersion relation as,
\begin{equation}
	\omega \qty(1 + 2\gamma_{_{EM}}^2 \omega^2 u_0^2 \, e^{2i\qty(\omega t - ksa)}) = \pm \sqrt{4\beta}\sin(\frac{ka}{2})
\end{equation}
We observe that the modification to the dispersion relation is depending on time and amplitude of the oscillation. In the limit $\gamma_{_{EM}}^2 \to 0$ we get the usual dispersion relation of elastic wave.

In the Debye approximation  velocity of sound\footnote{Note that here the dimension of $v_s$ is velocity$\times$(mass)$^{1/2}$/position} $v_s$ in the solid is constant as supposed to be for a classical elastic continuum.  The usual very-low-frequency vibrations of a solid are its acoustic oscillations and hence the dispersion relation at the low frequency limit  is written as
\begin{equation}
	\omega + 2\gamma_{_{EM}}^2 u_0^2 \omega^3 \qty[\qty(\cos(2\omega t) + 2kna \sin(2\omega t)) + i \qty(\sin(2\omega t) - 2kna \cos(2\omega t))] = \pm v_s k
\end{equation}
The term introduced due to \ac{gup} is complex and the elastic waves are influenced by the effect of gravity when traveling inside the solid, analogous to the electromagnetic waves in an interactive medium.

For our calculation we consider the dispersion relation of $\order{\gamma_{_{EM}}^2 }$ at $t = 0$ and  is given as,
\begin{equation}
	\omega + 2\gamma_{_{EM}}^2 u_0^2 \omega^3 = v_s k
\end{equation}
Thus, the \ac{gup} modified number of modes and accordingly the \ac{gup} modified density of states is given as,
\begin{equation}\label{eq:33}
	N^{^{GUP}} = \frac{V}{2\pi^2 v_s^3} \int \limits_0^{\omega_D} \dd{\omega} \omega^2 \qty(1 + 10\gamma_{_{EM}}^2 u_0^2 \omega^2) = \int \limits_0^{\omega_D} \dd{\omega} D(\omega)^{^{GUP}} = \frac{V\omega_{_D}^3}{6\pi^2v_s^3} \qty(1 + 6 \gamma_{_{EM}}^2 u_0^2 \omega_{_D}^2)
\end{equation}
Now, substituting \cref{eq:33,eq:15} in \cref{eq:25} the \ac{gup} modified total internal energy of the solid is given as\footnote{Here we have ignored the temperature independent terms of \cref{eq:15} for our convenience because thermodynamic properties are independent of it and therefore, it has no contribution to the specific heat.},
\begin{equation}
	U^{^{GUP}} = U + U\,'
\end{equation}
where,
\begin{subequations}
	\begin{align}
		U    & = 3\int\limits_0^{\omega_{_D}} \dd{\omega} \frac{V \hbar}{2\pi^2 v_s^3}\, \frac{\omega^3}{e^{\hbar \omega/ k_{_B} T} - 1}                                 \\
		U\,' & = 3\int\limits_0^{\omega_{_D}} \dd{\omega} \frac{V \hbar}{2\pi^2 v_s^3} \gamma_{_{EM}}^2 \qty[10 u_0^2 \, \frac{\omega^5}{e^{\hbar \omega/ k_{_B} T} - 1}
			+ \hbar \sum\limits_{j=1}^\infty \qty(1 - j \frac{\hbar \omega}{k_{_B} T}) j \omega^4 e^{-j\hbar \omega/ k_{_B} T}]
	\end{align}
\end{subequations}
Finally, we calculate the \ac{gup} modified Debye specific heat as,
\begin{equation}
	C_v^{^{GUP}} = \dv{U^{^{GUP}}}{T} = C_{v} + C\,'_{v}
\end{equation}
where,
\begin{subequations}
	\begin{align}
		C_{v}    & = \int\limits_0^{y_{_D}} \dd{y} 9N k_{_B} \frac{T^3}{\theta_{_D}^3} \frac{y^4 e^y}{\qty(e^y - 1)^2}                                                                                              \\
		C\,'_{v} & = \int\limits_0^{y_{_D}} \dd{y} 9Nk_{_B} \gamma_{_{EM}}^2 \qty[ u_0^2 \frac{k_{_B}^2}{\hbar^2}  \frac{T^3}{\theta_{_D}}\qty(-6 + 10 \frac{T^2}{\theta_{_D}^2}y^2)\frac{y^4 e^y}{\qty(e^y - 1)^2}
			+ k_{_B} \frac{T^4}{\theta_{_D}^3} y^5 \sum\limits_{j=1}^\infty\qty(2 - yj) j^2 e^{-jy} ]
	\end{align}
\end{subequations}
where\qcomma $\theta_{_D} = \hbar \omega_{_D} / k_{_B}$ is the characteristic temperature known as Debye's temperature, $y = \hbar \omega / k_{_B} T$ and $y_{_D} = \theta_{_D} / T$. In the limit $\gamma_{_{EM}}^2 \to 0$ we get back our standard Debye's specific heat.

\begin{figure}[ht]
	\centering
	\includegraphics[scale=0.4]{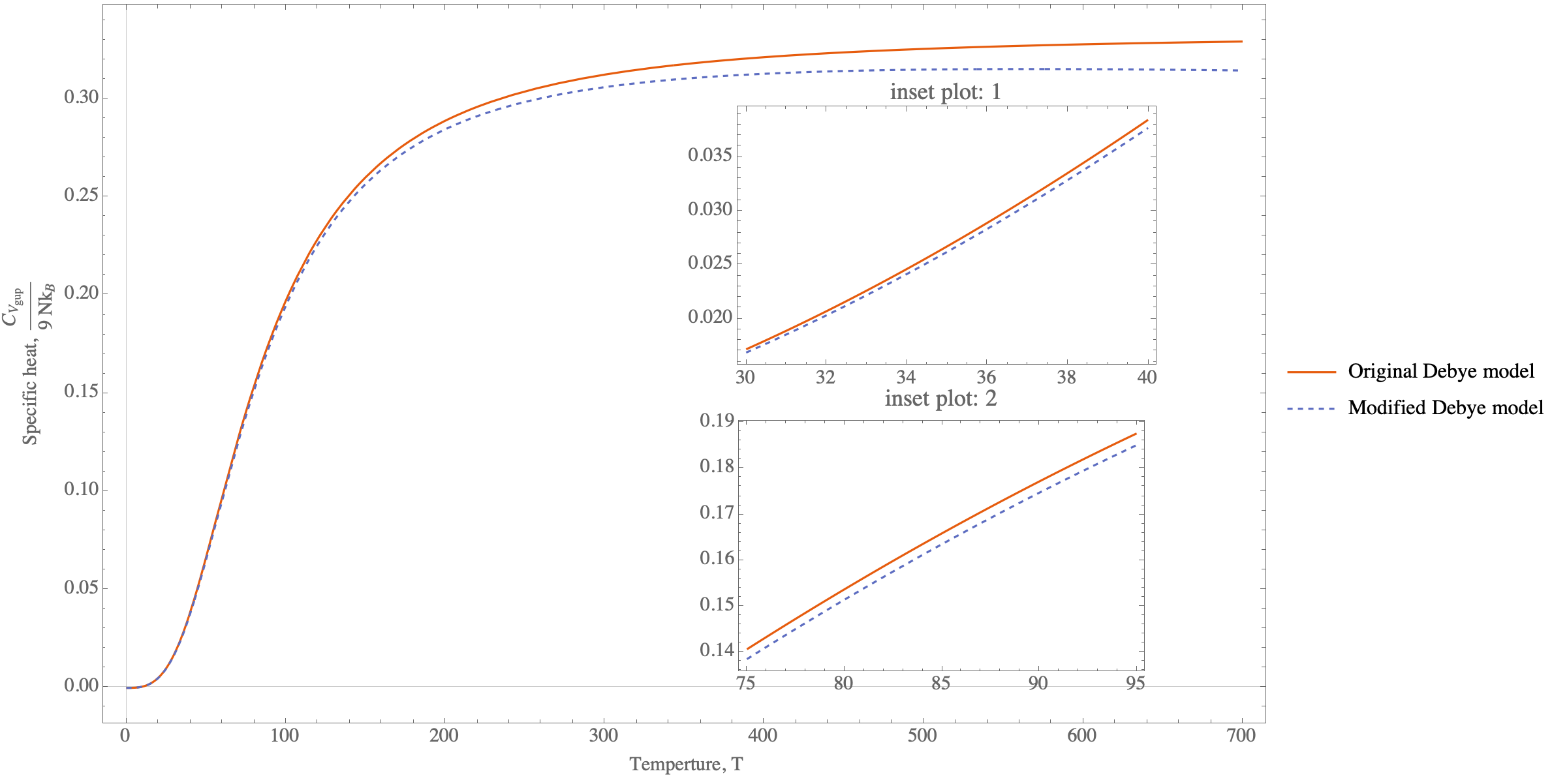}
	\caption[modified Debye]{Modified Debye specific heat, $C_{v_{GUP}} / 9Nk_{_B}$ vs. Temperature, $T$ plot. Here we consider $\theta_D \approxeq 343K$ (for copper \cite{mahmood-Experimental-2011}) and the range of temperature $0K \leq T\leq 700K$. We have chosen the \ac{gup} dependent parameter $\gamma_{_{EM}}^2 u_0^2 k_{_B}^2 / \hbar^2 = 10^{-45}$ and $\gamma_{_{EM}}^2 k_{_B} = 10^{-4.5}$. Inset plot $1$ and $2$ shows the variation of specific heat at $30K \leq T\leq 40K$ and $75K \leq T\leq 95K$ respectively.}
	\label{fig:3}
\end{figure}

We now study the following cases,
\paragraph{Behavior at high temperature:} For temperatures $T \gg \theta_{_D}$ or $y_{_D} \ll 1$, $y$ will be very small, and we can write $e^y \approxeq 1 + y + \order{y^2}$ and $e^{-jy} \approxeq 1 -j y + \order{y^2}$. Thus, at high temperature the \ac{gup} modified specific heat of the solid is given as
\begin{equation}
	C_v^{^{GUP^{\uparrow}}} = C_{v}^{^\uparrow} + C_{v}'{^{^\uparrow}}
\end{equation}
where,
\begin{subequations}
	\begin{align}
		C_{v}^{^\uparrow}    & = 3Nk_{_B}      \label{eq:39a}                                                          \\
		C_{v}'{^{^\uparrow}} & = 9Nk_{_B} \gamma_{_{EM}}^2 \qty[u_0^2 \frac{k_{_B}^2}{\hbar^2} \frac{\theta_{_D}^3}{T}
			+   k_{_B} \frac{\theta_{_D}^3}{T^2} \sum\limits_{j=1}^\infty\qty(\frac{1}{3} - \frac{3j}{7} \frac{\theta_{_D}}{T})j^2] \label{eq:39b}
	\end{align}
\end{subequations}
The original term \cref{eq:39a} gives us the usual value obtained from classical theory. But, the summation in \cref{eq:39b} \ie, the additional term due to \ac{gup} have a negative value. It also has a $1/T$ dependence at high temperatures. Therefore, the overall Debye specific heat drops from the usual value as observed in \cref{fig:3}.

\paragraph{Behavior at low temperature:} For temperatures $T \ll \theta_{_D}$ or $y \gg 1$ the value of $y_{_D}$ shoots up to infinity. Thus, at low temperature the \ac{gup} modified specific heat of the solid is given as
\begin{equation}
	C_v^{^{GUP^{\downarrow}}} = C_{v}^{^\downarrow} + C_{v}'{^{^\downarrow}}
\end{equation}
where,
\begin{subequations}
	\begin{align}
		C_{v}^{^\downarrow}    & = 12Nk_{_B} \pi^4 \frac{T^3}{\theta_{_D}}   \label{eq:42a}                                                  \\
		C_{v}'{^{^\downarrow}} & = - 72Nk_{_B} \gamma_{_{EM}}^2  \qty[u_0^2 \frac{k_{_B}^2}{\hbar^2} \frac{\pi^4}{5} \frac{T^3}{\theta_{_D}}
			+  k_{_B} \frac{2 \pi^4}{3} \frac{T^4}{\theta_{_D}^3}]\label{eq:42b}
	\end{align}
\end{subequations}
Here both the original term \cref{eq:42a} and the additional term introduced due to \ac{gup} \cref{eq:42b} is $T^3$ dependent. But the overall specific heat of the solid at low temperatures will stay at a lower value than usual as observed in \cref{fig:3} because of the negative value of the \ac{gup} introduced term. In addition to this, we can see a $T^4$ dependence of the \ac{gup} term at low temperature.

\begin{figure}[ht]
	\centering
	\includegraphics[scale=0.4]{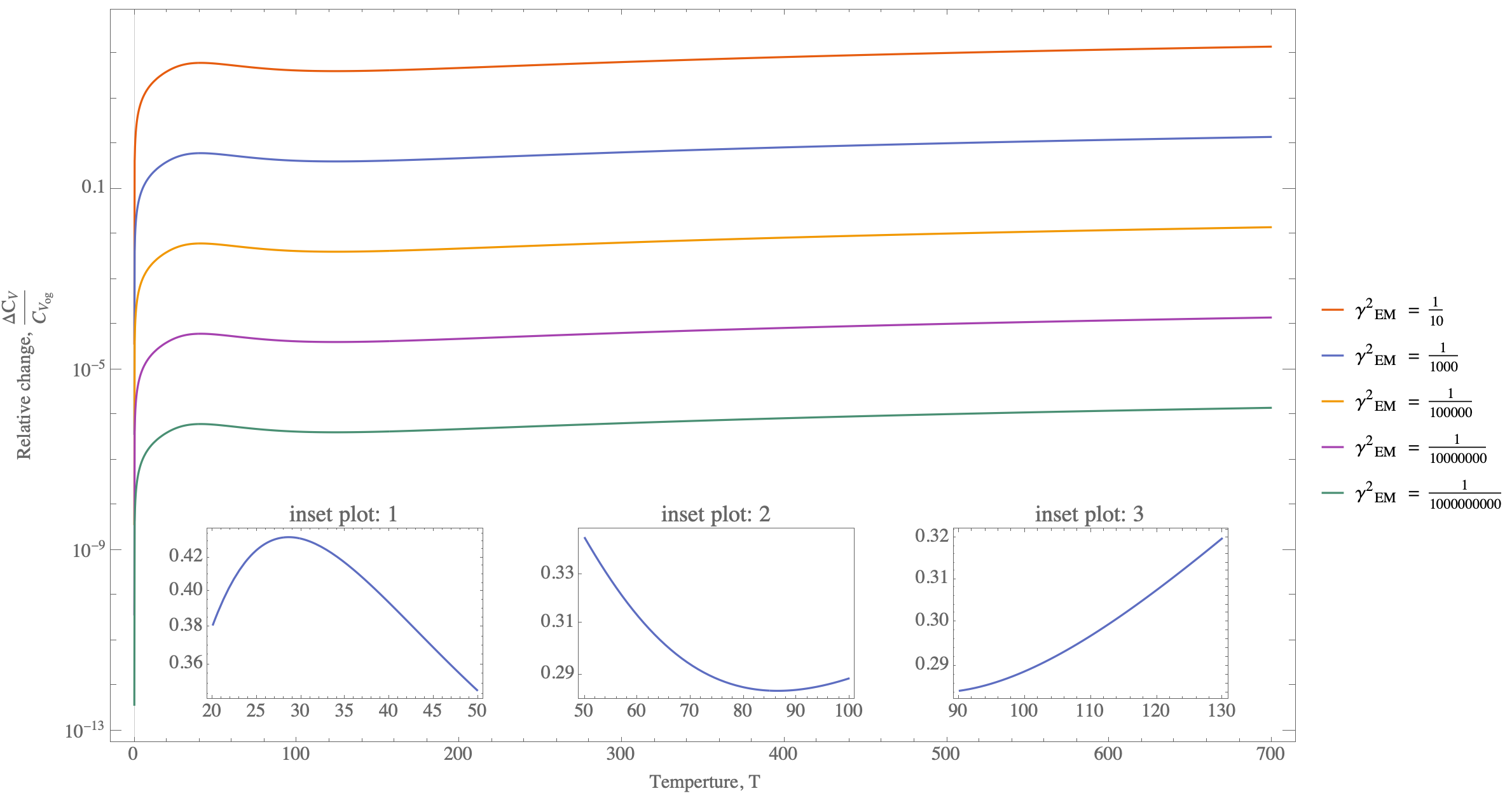}
	\caption[relative Debye]{Relative Debye specific heat, $\Delta C_v/ C_{v}$ vs. Temperature, $T$ plot at $\theta_D \approxeq 343K$ with temperature ranging from  $0K$ to $700K$. We vary the \ac{gup} parameter as $\gamma_{_{EM}}^2 = \qty{10^{-1}, 10^{-3}, 10^{-5}, 10^{-7}, 10^{-9}}$. Inset plot $1,2$ and $3$ shows the relative change of specific heat at $20K\leq T \leq 50K$, $50K\leq T \leq 100K$ and $90K\leq T \leq 130K$ respectively for the \ac{gup} parameter $\gamma_{_{EM}}^2 = 10^{-3}$}
	\label{fig:4}
\end{figure}

In \cref{fig:4} we observe the relative modification of the Debye specific heat model using \cref{eq:24}. The difference between original specific heat and the \ac{gup} modified specific heat falls rapidly as the \ac{gup} parameter $\gamma_{_{EM}}^2$ gets smaller. And for a particular value of $\gamma_{_{EM}}^2$, the difference rises below $30K$ and then starts to fall moderately from $30K$ to $85K$. Beyond $90K$ a continuous growth in the difference is observed.

\section{Conclusion}\label{sec:4}
In this paper we have studied the influence of quantum gravity theories on specific heat of solids using both Einstein as well as Debye model. We obtain the modifications  to the specific heat in Einstein's model using the GUP modified spectrum of QHO. Our results match with the usual expression in the limit $\gamma_{_{EM}}^2 \to 0$. The resulting specific heat is found to be less than the standard value at any temperature. In the case of  Debye model of  specific heat  we  have checked that the propagation speed of waves in the solid is  influenced by the gravity. We calculate  the modification  to the dispersion relation which  depends on amplitude and keeps growing with time. It is also noted that the modification is complex suggesting that some kind of interaction is happening analogous to the electromagnetic waves in an interactive medium. Furthermore, we calculate the modification to the specific heat in Debye's model which matches the standard value when we ignore the quantum gravity effects. At low temperature, the modification is negative and receives contributions proportional to   both  $T^3 $ as well as $T^4$. The departure of the heat capacity from the classical value (Dulong-Petit law) at low temperatures is one example of the success of quantum mechanics in describing experimental observations. Thus, the observations here may be used to determine the effect of gravity in the specific heat models of a solid experimentally. However GUP modifications are extremely small. It will be interesting  to study the \ac{gup} modified dispersion relation considering the time dependence in future.

\section*{Acknowledgement}\label{sec:5} One of us (BPM) acknowledges the research grant for faculty under IoE Scheme (Number 6031) of Banaras Hindu University, Varanasi.
\bibliographystyle{unsrtnat}
\bibliography{citation.bib}
\end{document}